\documentclass[pra,twocolumn,superscriptaddress]{revtex4}
\usepackage{color}
\usepackage{amsfonts}
\usepackage{amsmath}
\usepackage{amssymb}
\usepackage{graphicx}

\usepackage[usenames,dvipsnames]{xcolor}
\usepackage{soul}

\begin{document}

\title{Shot noise in Weyl semimetals}

%\author{P.G. Matveeva$^{1}$,  D.N. Aristov$^{1}$, D.  Meidan$^{2}$  and  D.B. Gutman$^{3}$}
\author{P.G. Matveeva}
\affiliation{Department of Physics and Research Center Optimas, University of Kaiserslautern, 67663 Kaiserslautern, Germany}
\affiliation{``PNPI'' NRC ``Kurchatov Institute'', Gatchina 188300, Russia}
\author{D.N. Aristov}
\affiliation{``PNPI'' NRC ``Kurchatov Institute'', Gatchina 188300, Russia}
\affiliation{Department of Physics, St.Petersburg State University, Ulianovskaya 1,
St.Petersburg 198504, Russia}
\affiliation{Institute for Nanotechnology, Karlsruhe Institute of Technology, 76021
Karlsruhe, Germany }
%\affiliation{$^1$ Department of Physics, St.Petersburg State University, Ulianovskaya 1, St.Petersburg 198504, Russia}
\author{D.  Meidan}
\affiliation{Department of Physics, Ben-Gurion University of the Negev, Beer-Sheva 84105, Israel}
\author{D.B. Gutman}
\affiliation{Department of Physics, Bar-Ilan University, Ramat Gan, 52900, Israel}

\begin{abstract}
We study the effect of inelastic processes on the magneto-transport of a quasi-one dimensional Weyl semi-metal, using a modified Boltzmann-Langevin approach. The magnetic field drives a crossover to a ballistic regime in which the propagation along the wire is dominated by the chiral anomaly, and  the  role of  fluctuations inside the sample is exponentially suppressed. We show that inelastic collisions modify the  parametric dependence of the current fluctuations on the magnetic field. By measuring shot noise as a function of  a magnetic field, for different applied voltage, one can estimate the   electron-electron inelastic  length $l_{\rm ee}$.
 %We show that the shot noise is a universal function of electron-electron inelastic scattering, sample size and a drift length.  
\end{abstract}
\date{\today}
\preprint{}

\maketitle

\section{Introduction} 
Weyl semimetals have been a subject of  active experimental and theoretical research due to their unusual transport properties \cite{Turner,Wan,Burkov_2011,Hosur,Kim,Xu,Weng,Huang,Neupane,Liu,Xu_2015,Weng_2015}. 
A Weyl semimetal is characterized by a three dimensional  band structure, where valence and conduction band touch at discrete  isolated points in the Brillouin  zone. Excitations in the vicinity of the band degeneracy points are governed by the Weyl Hamiltonian $H=\pm \hbar\, \mathbf{ p}\cdot\bold{\sigma}$. 

The non-trivial topology of Weyl semimetal can be revealed by applying  an external magnetic field, which leads to the formation of  Landau levels (LL).  
The  Weyl nodes result in the emergence of chiral zero Landau levels,  protected against scattering for a sufficiently smooth disorder. 
Remarkably, transport properties in the presence of a magnetic field can be described by the semiclassical Boltzmann equation\cite{Burkov_2014,Son2013,Para2014}.   
\begin{eqnarray}
\label{Kinetic}
\frac{\partial f(\bold{p},{\bf r})}{\partial t} + \dot{\bold{r}} \frac{\partial f(\bold{p},{\bf r})}{\partial r} + \dot{\bold{p}} \frac{\partial f(\bold{p},{\bf r})}{\partial \bold{p}} = I[f(\bold{p},{\bf r})] .
\end{eqnarray} 
Here $f(\bold{p}) $ is the distribution function, $\bold{p} $ is the momentum, $ I[f(\bold{p})] $ is the collision integral and:
 \begin{equation}
\begin{aligned}
\dot{\bold{r}} &= \frac{\partial \epsilon_p}{\partial \bold{p}}+\dot{\bold{p}}\times\bold{\Omega}_\mathbf{p}  \,,\\
\dot{\bold{p}} &= e{\bf E}+\dot{\bold{r}}\times e\bold{B} \,.
\end{aligned}
\label{rdot}
\end{equation}
The nontrivial topology is reflected in  Berry curvature terms that appear  in the Liouville operator \cite{Niu,Xiao2010,Nagaosa2010}:
 \begin{equation}
\begin{aligned}
\bold{\Omega}_{\mathbf{p}} &= \bold{\nabla}_\mathbf{p}\times \bold{A}_\mathbf{p}\,,\\
 \bold{A}_\mathbf{p} &= i\langle u_\bold{p} |  \bold{\nabla}_\mathbf{p} u_\bold{p} \rangle\,,
\end{aligned}
%\label{}
\end{equation}
where $u_\bold{p}$ is a periodic part of the Bloch wave function.
These terms  originate from the chiral anomaly\cite{Nielsen1983} and are absent in topologically trivial matter. They give  rise to ballistic  propagation\cite{Para2014}  and  non-local ac conductance\cite{Stern2015}.

In this work, we study the effect of inelastic processes on the magneto-transport of a quasi-one dimensional Weyl semi-metal, within a  semi-classical  description.  The wire,  with  cross-section $A$ is pierced by  a magnetic field $B$,  oriented along the wire in $\hat{z}$ direction.
We assume that the inter-nodal scattering length is much longer than the system length $L$.
This  allows us to focus in the vicinity of a single Weyl node with a given chirality. The final answer is given by a sum over all Weyl nodes. 

We first note that similar to a metal, the conductance is unaffected by interactions between electrons, as those do not lead to momentum relaxation. 
Moreover, as the conductance is measured as a response to the difference of total electro-chemical potentials, the results obtained for interacting electrons within self-consistent field approximation, 
and for non-interacting electrons subjected to a difference in chemical potential, coincide. 
Consequently, the conductance can be calculated in the limit of non-interacting Weyl fermions, neglecting both the inelastic collision integral and the self-consistent electric field.  

Using Eq. \eqref{rdot} for non-interacting Weyl fermions of positive chirality, the density of electric current is given by: 
\begin{eqnarray}\label{current_density}
\bold{j}({\bf r}) =  \int \frac{d^3\mathbf{p}}{(2\pi\hbar)^3}\left[ \frac{\partial \epsilon_p}{\partial \bold{p}}+\frac{e{\bf B}}{c}\bold{\Omega}_p\cdot\frac{\partial \epsilon_p}{\partial \bold{p}}\right]  f(\bold{p},{\bf r}).
\end{eqnarray}
Assuming that  intra-nodal scattering is short ($l_{\rm intra} \ll L$), 
we perform a diffusive approximation, %\newline
$f(\bold{p},{\bf r})=  f(\epsilon_{p},z) +(\hat{z} \cdot \bold{p} ) \frac{\tau}{m} \frac{\partial  f(\epsilon_{p},z) }{\partial z}$, where $\tau$ is a momentum relaxation (transport) time, $m$ is an effective mass of an electron.
The current density within diffusion approximation is 
\begin{equation}
\label{av_current}
\bold{j}(z)=\left(D\partial_z\rho - \frac{eB}{4\pi^2\nu}\rho\right),
\end{equation} 
where the density of electrons
\begin{equation}
\label{density}
\rho(z)=\nu\int_{-\infty}^\infty d\epsilon\,  f(\epsilon,z).
\end{equation}
Using the fact that the current density is subject to the continuity equation, which in the static limit reads $\partial_z \bold{j}(z)=0$,
we find that the electron density obeys a  drift-diffusion equation \cite{Burkov_2014,Son2013,Para2014}  characteristic of a one dimensional  random walk with  
non-equal probabilities:
\begin{equation}
\label{Drift_Diffusion}
\xi \,\partial_z^2\rho(z) \mp N_{\phi}\partial_z\rho(z)=0 \,. 
\end{equation}
Here $\mp$ signs correspond to Weyl nodes of opposite chiralities,  $\xi$ is the localization length ($\xi=2\pi \nu D A$, with  $\nu$   three  dimensional  density of states and $D$   diffusion constant),  and
\begin{equation}
N_\phi=\frac{AeB}{\Phi_0}=\frac{A}{2\pi} \frac{eB}{\hbar c} \, .
\end{equation}
is the number of magnetic flux quanta  piercing the wire, which also marks the imbalance between left and right moving modes.

The validity of  Eqs. \eqref{av_current} and \eqref{Drift_Diffusion} was recently established with Keldysh \cite{Bagrets} and super-symmetry \cite{Khalaf2016,Khalaf2017} non-linear sigma model formalism.  
Eq. \eqref{Drift_Diffusion} shows that while  the disorder scattering between different LL  disrupts  the ballistic propagation of the  zero Landau level, the imbalance between the number of left and right moving modes (for a given Weyl node) remains. 
 Eq.  \eqref{Drift_Diffusion} should be supplemented with  the boundary conditions for the value of distribution function at the leads. For the sample subject to voltage difference, the boundary conditions are
 $f(z=0)=f_F(\epsilon-eV/2), f(z=L)=f_F(\epsilon+eV/2)$; here 
$f_F(\epsilon)$ is  Fermi-Dirac distribution function.  

%$f(\epsilon,z)$  obeys a drift-diffusion equation \cite{Burkov_2014,Son2013,Para2014}  characteristic of a one dimensional random random walk with  
%non equal probabilities:
%\Dadd{\begin{equation}
%\label{Boltzmann}
%\xi \partial_z^2 f(\epsilon,z) \mp N_{\phi}\partial_z f(\epsilon,z)-I_{\rm e-e}[f]-I_{\rm e-ph}[f]=0
%\end{equation}
%Here $\mp$ signs correspond to Weyl nodes of opposite chiralities,  $\xi$ is the localization length ($\xi=2\pi \nu D A$,   $\nu$ is a three  dimensional  density of states and $D$ is a diffusion constant), 
%\begin{equation}
%N_\phi=\frac{AeB}{\Phi_0}=\frac{A}{2\pi} \frac{eB}{\hbar c}.
%\end{equation}
%is the number of magnetic flux quanta  penetrating the wire, and
%$I_{\rm  e-e}$  and $I_{\rm  e-ph}$  are electron-electron and electron-phonon collision  integrals. We note that $N_\phi $ marks the imbalance between left and right moving modes.}
%The validity of  Eqs. \eqref{av_current} and \eqref{Boltzmann} was recently established with Keldysh \cite{Bagrets} non-linear sigma model formalism.  
%Their  physical interpretation is as follows.
%Disorder scattering between different LL  disrupts  the ballistic propagation of the   zero Landau level. 
% However, the imbalance between the number of left and right moving modes (for a given Weyl node) remains. 
% As a result of this imbalance,  

%Solving for the distribution function $ f(\epsilon_p)$, the current density is then given by:
%which within  the diffusion approximation becomes:
Solving Eq. (\ref{Drift_Diffusion}) for the electron density and using Eq. \eqref{av_current} we obtain the conductance
\begin{equation}
G(B)=N_W\frac{e^2 N_\phi}{4\pi\hbar}\coth\left(\frac{L}{2a}\right)\,,
\label{GB}
\end{equation} 
in agreement with  Keldysh \cite{Bagrets} and  super-symmetry sigma model calculations \cite{Khalaf2016,Khalaf2017}. 
Here the drift length  $a \equiv \xi/N_\phi=   2\pi\nu D/eB$, and we have multiplied the result by the total number of Weyl nodes $ N_W$.
The semiclassical description \eqref{Drift_Diffusion} shows that the magnetic field drives a crossover from diffusive propagation to a ballistic regime, where only the chiral channels contribute to transport, leading  in particular,   to a positive magneto-conductance \cite{Nielsen1983,Fukushima2008,Zyuzin2012,Aji2012}.

%As in normal metals, inelastic scattering do not affect the conductance, as these do not lead to momentum relaxation, and the result follows the noninteracting limit \cite{Bagrets,Khalaf2017}.

We note that while the  semiclassical approximation restricts the strength of   magnetic fields $\omega_c\tau \ll 1$, the value of $ L/a$ may be large provided that  $L >E_F\tau l$, where $l$ is  an elastic  mean free path and $E_F$ is Fermi velocity.
Moreover, in order to  ignore localization effects,  the sample should be shorter than   localization length $ L/\xi\ll1$.
Both conditions can be fulfilled simultaneously, provided  $Ap_F^2 \gg E_F\tau$.

%\begin{equation}
%\label{Boltzmann}
%\xi \partial_z^2 f(\epsilon,z) \mp N_{\phi}\partial_z f(\epsilon,z)-I_{\rm e-e}[f]-I_{\rm e-ph}[f]=0.
%\end{equation}

%\st{To be able to use semiclassical approach in the presence of Landau quantization, one assumes 
%that the magnetic field is sufficiently weak.
%If   the inter and intra LL scattering rate are equal, the LL are mixed for
%$\omega_c\tau \ll 1$, where $\omega_c$ is a cyclotron frequency and $\tau$ is elastic scattering time.
%For a generic disorder  it the inter and intra LL may be significantly different.
%For example, for   $z$ independent  disorder potential $V_{\rm dis}(x,y)$ there is no scattering  between states that belong to different LL. 
%In this case, the propagation of electrons on zero LL is entirely ballistic, 
%and diffusive for all other LL.
%However, for a generic case, there are finite (not necessarily equal) inter and intra LL scattering times, 
%and corresponding mean free paths.  We will be interested in  systems that are much longer that elastic scattering length
%relatively to both,  inter and intra LL scattering length.  }

\section{Fluctuations in disordered Weyl semimetals}
We now turn to a computation of current noise. 
When calculating the low frequency noise, the self-consistent electric field can be neglected. This is because it 
 leads to the replacement of the inverse Thouless time  with the inverse Maxwell time as a characteristic frequency below which the zero frequency limit is achieved.
The inelastic collisions on the other hand play an important role for non-equilibrium current fluctuations, and the corresponding collision integrals needs to be restored in the kinetic equation \eqref{Kinetic}. Within  the diffusive approximation, the latter reads:
\begin{equation}
\label{Boltzmann}
\xi \partial_z^2 f(\epsilon,z) \mp N_{\phi}\partial_z f(\epsilon,z)-I_{\rm e-e}[f]-I_{\rm e-ph}[f]=0 \,,
\end{equation}
%Here $\mp$ signs correspond to Weyl nodes of opposite chiralities,  $\xi$ is the localization length ($\xi=2\pi \nu D A$,   $\nu$ is a three  dimensional  density of states and $D$ is a diffusion constant), 
%\begin{equation}
%N_\phi=\frac{AeB}{\Phi_0}=\frac{A}{2\pi} \frac{eB}{\hbar c}.
%\end{equation}
%is the number of magnetic flux quanta  penetrating the wire, and
where $I_{\rm  e-e}$  and $I_{\rm  e-ph}$  are electron-electron and electron-phonon collision  integrals.

For  topologically trivial metals,   current fluctuations  can be computed following a   Boltzmann-Langevin approach\cite{Kogan,Shulman-Kogan}. This approach  relates the fluctuation of observable quantities, such as current, to the fluctuation of the occupation in the phase space
 $\delta f({\bf p},{\bf r},t)$, with the same applicability as  the kinetic equation. 
We  employ this approach,   taking into account effects of topology present in Weyl semimetals.  

Similar to the current density \eqref{current_density}, the density of current fluctuations in a Weyl semi-metal  can be related to the fluctuation of phase space density as  
\begin{equation}
\label{deltaj}
 \delta {\bf j}(t)=\sum_{\bf p}\bigg[ \frac{\partial \epsilon_{\bf p}}{\partial  {\bf p}}\delta f({\bf p},{\bf r},t)+\frac{eB}{4\pi^2\nu}\delta f({\bf p},{\bf r},t)\bigg].
\end{equation}
The first term in Eq. \eqref{deltaj}  is affected by the antisymmetric part of the momentum-dependent distribution  function, while the second term is proportional to its symmetric part, i.e. the total density of electrons at a given point. This second term is absent in topologically trivial metals and corresponds to the drift current associated with electrons occupying the zero Landau level.

%At equilibrium the current fluctuations  are related to linear response via Fluctuation-Dissipation theorem (FDT)
%\begin{equation}
%S_2(\omega) \equiv \langle \delta I(t) \delta I(0) \rangle
%=\omega \coth \left(\frac{\omega}{2T}\right) G(\omega).
%\end{equation}
%Though we are aiming at non-equilibrium case, Boltzmann-Langevin approach should reproduce     
 %the low frequency limit ($\omega \ll T$) 
 %We now proceed with calculation of the second moment of current fluctuations.

We note that due to particle number conservation, the  correlation function of current fluctuation is independent of the lateral  position $z$. This allows  to compute the correlation of integrated currents: 
\begin{equation}
\delta I(t)=\frac{1}{L}\int d^3 \mathbf{r}\, \delta{\bf  j}({\bf r} ,t)\,.
\end{equation} 

The  correlation function of current fluctuation is calculated by representing the  kinetic theory of fluctuations  through a Keldysh field theory\cite{Kamenev_book}.
This approach  is identical to the original Boltzmann-Langevin equation for pair-correlation function but allows to compute current  cumulants of any order. 
While our main focus in this manuscript is on the pair-correlation function only, the calculation presented hereafter enables one to  obtain the full kinetic theory of fluctuations in Weyl semimetals.

The generating function of counting statistics\cite{Levitov}  can be written as \cite{Gutman2004,Pilgram,Derrida}
\begin{equation}\label{PI}
\kappa[\lambda] = \int\!
Df \,D\delta\bar{f} \,  e^{i{\cal S}[f,\delta\bar{f},\lambda]+i \int\! dt dz \lambda(t)\delta I(z,t)}\,.
%Df({\bf p}, {\bf r}, t)D\delta\bar{f} ({\bf p}, {\bf r}, t)e^{i{\cal S}[f,\delta\bar{f}]+i \int\! dt dz \lambda(t)\delta I(z,t)}\,.
\end{equation} 
Here the density of particles in the phase space is treated as a field $f\equiv f({\bf p}, {\bf r} , t)$, and the  field $\delta\bar{f}\equiv \delta\bar{f}({\bf p}, {\bf r} , t)$ is its conjugate. The field $f({\bf p}, {\bf r} , t)$ consists of a mean value, which solves the Boltzmann equation \eqref{Boltzmann}, and a fluctuation part $\delta f({\bf p},{\bf r},t)$.
The auxiliary counting field,  $\lambda$,  is chosen in accordance with the  correlation function in question.
The effective action consists of two parts
\begin{equation}
\label{action}
{\cal S}[f,\delta\bar{f},\lambda] = {\cal S}_{\rm Dyn} + {\cal S}_{\rm Noise}[\lambda] \,.
\end{equation} 
The dynamical part of the action
\begin{widetext}
\begin{equation}\label{S_dyn}
i{\cal S}_{\rm Dyn} = 2\pi i \nu A \int dt \,d\epsilon \, dz\bigg[ 
\delta \bar{f}(\epsilon,z,t)
\left\{
\frac{\partial}{\partial t}-D \frac{\partial^2 }{\partial z^2}+\frac{eB}{4\pi^2\nu}\frac{\partial}{\partial z}+\hat{I}_{\rm e-ph}+\hat{I}_{\rm e-e}
\right\}f(\epsilon,z,t)
\bigg]\,, 
\end{equation}
\end{widetext}
describes the evolution of the particle occupation in the phase space.
The cascade  creation \cite{Nagaev2002}  of the noise is encoded in  
\begin{eqnarray}
i{\cal S}_{\rm Noise}[\lambda]&=&-\nu A \int dt\, d\epsilon\, dz \left(\lambda +\frac{\partial \delta \bar{f}(\epsilon,z,t)}{\partial z} \right)^2 \nonumber \\
& \times & f(\epsilon,z,t)(1-f(\epsilon,z,t)).
\end{eqnarray}

There are two modifications of the action in Eq. \eqref{action}  as compared  with the one for  normal metals \cite{Gutman2004,Pilgram,Derrida}.
First, the third term (in curly brackets) in Eq. \eqref{S_dyn} is absent in normal metals. This term dictates   that the  dynamics of noise propagation on scales longer than the elastic collision length  is of drift-diffusion type, as opposed to diffusion in normal metals.
The second is more subtle and is related to the fluctuating current  that couples to the source term in Eq. \eqref{PI}. In normal metals $ \delta \mathbf{j}$ is proportional to the velocity $ {\bf v_p} $ (see the  first term in Eq. \eqref{deltaj}), which is related to spatial gradients of the density within the diffusive approximation. For this reason,  in topologically trivial metals it contributes to the mean value of the current only which is determined by gradients of the density, see Eq. \eqref{av_current}, and plays no role in its fluctuations, as the source term in Eq. \eqref{PI} couples to the integrated current.  In Weyl semimetals, on the other hand,   this term  has a drift  contribution, encoded in the second term in Eq. \eqref{deltaj}, which  has no gradients. For that reason it contributes to the  fluctuations of currents.
 
Using the  action (\ref{action})  one expresses the current fluctuations
\begin{equation} 
S_2=\int dt \,\langle \delta I(t) \delta I(0)\rangle
\end{equation}
through the correlation functions of phase space density  operators
\begin{widetext}
\begin{eqnarray}&&
\label{gen_noise}
S_2=\frac{e^2}{L^2}\int d1d2 \bigg\{
 \delta(1-2) f(1)(1-f(1))- \frac{\xi^2}{2}
\bigg[ 
a^{-2}D^K(1,2)+4a^{-1}D^R(1,2)\partial_{z_2}f(2)(1-f(2))
\bigg]
\bigg\}.
\end{eqnarray}
\end{widetext}
Where we have used  $1\leftrightarrow (\epsilon_1,z_1)$, $d1 \leftrightarrow d\epsilon_1dz_1$ and similarly for $2$, for brevity; 
 $D^{K/R}$ are Keldysh and retarded components of $\delta f$ correlation functions.
Computing the correlation functions, in the presence of inelastic scattering (see  Appendix \ref{derivation} for the details)
one finds the low-frequency noise spectrum 
\begin{equation}
\label{s2}
S_2=
\frac{e^2 \xi}{a^2}
\int_0^L dz \,T(z) \frac{\exp(2(L-z)/a)}{(e^{L/a}-1)^2}\,.
\end{equation}  
The   information about non-equilibrium state of the system is encoded in  effective temperature
\begin{equation}
T(z)=\int_{-\infty}^\infty d\epsilon\,  f(\epsilon,z)(1-f(\epsilon,z)) \,,
\end{equation}
which accounts for the spread of  the distribution function $f(\epsilon,z) $ determined by Eq.(\ref{Boltzmann}) in the presence of all scattering processes.

Eq.(\ref{s2}) is the central  result of this work.   It extends the standard expression for the noise spectrum in normal metals 
to Weyl semimetal. While in  normal metals the noise is proportional to an integral over effective temperature alone, for a Weyl semimetal it acquires an additional kernel, 
which depends exponentially on the lateral distance from the leads.
The physical reason  for that  has to do with the chiral anomaly.  In normal matter, fluctuations  of phase space distribution that occur anywhere in the sample  diffusively propagate 
to the contacts, giving rise to a current noise.
For Weyl semimetal, on the other hand, on  scales much greater than $a$, the  fluctuations  in the  phase-space occupation  move  predominantly in a ballistic  fashion. 
Hence,  in longe samples,  $L\gg a $, where the transport is dominated by ballistic (deterministic) propagation, stochastic processes are exponentially suppressed, and the system is noiseless.
 % \Dcom{Isnt it that the important effect of chirality is the introduction of balistic motion at $L\gg a$? In which case the reason for the exponential has to do with the lack of randomization rather than the directionality. In other words, Drift motion which is also associated with chiral modes is not noiseless.}
 
 We now analyze the noise  for a number of limiting cases and 
start  from the limit  of large inelastic length. 

\subsection{Elastic scattering}
At equilibrium, the distribution function is of Fermi-Dirac form  and  Eq. (\ref{s2}) yields
\begin{equation}
S_2=2T \, G(B) \,,
\end{equation}
with $G(B)$ from \eqref{GB}, 
in accordance with the   fluctuation-dissipation theorem. % From now on we set $N_{W}=2$.
In fact, this result holds even in the presence of inelastic scattering, as can be easily checked by imposing thermal equilibrium, which implies  a uniform temperature $ T(z)=T$ in Eq.\ \eqref{s2}.

\begin{figure}
\includegraphics[width=0.9\columnwidth]{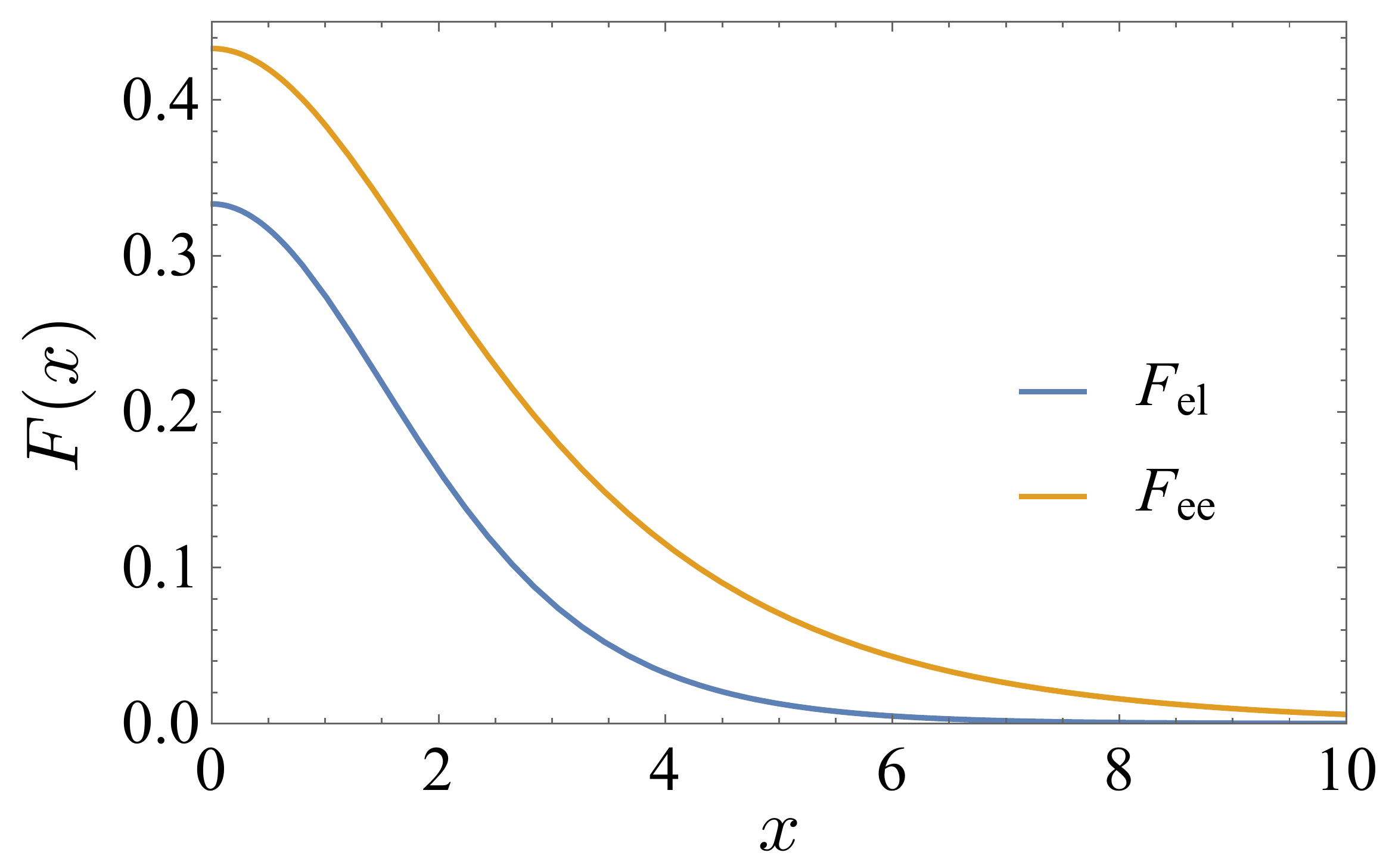}%{Fano_factor.pdf}
\caption{Fano factors for elastic ($F_{\rm el}$ lower) and inelastic ($F_{ee}$ upper) regimes as functions of the ratio, $x=L/a$, of the system length $L$ to the drift length $a =2\pi\nu D/eB $.}
\label{figure1}
\end{figure}

Next, we study the shot noise in  the limit $T=0$, and finite bias $V$.
Substituting Eqs.\ (\ref{el_diff}) for non-interacting electrons into Eq.\ (\ref{gen_noise}) one  arrives at Eq.\ (\ref{s2}) that yields
 \begin{equation}
\begin{aligned}
S_2 
%& 
%=\frac{e^3 N_{W} \xi V }{16\pi L}\frac{x\sinh x-x^2}{\sinh^4(x/2)} \,, \\ &
= \frac{e^3 N_{W} N_{\phi} V }{16\pi }\frac{\sinh x -x}{\sinh^4(x/2)} \,,
\end{aligned}
% \label{}
\end{equation}
with $x=L/a$ in agreement with \cite{Bagrets,Khalaf2017}.
It corresponds  to  Fano factor ($F=S_2/eI$)
\begin{equation}
\label{Fel}
F_{\rm el}(x)=\frac{\sinh{x}-x}{2\sinh{x}\sinh^2(x/2)}.
\end{equation}
At short distances $ L \ll a$, the Fano factor approaches the value  of diffusive metals, $1/3$, while it  is exponentially suppressed for $L \gg a$, see Fig.\ref{figure1}.
This suppression  indicates that  at large sample sizes (or large magnetic fields),  $L \gg a$, all electronic  motion is predominantly ballistic.

\subsection{Electron-phonon relaxation ($L \gg l_{\rm e-ph}$)}
\label{el_ph}
We now consider the case when the sample is much longer that electron-phonon inelastic length, yet momentum relaxation is still governed by  static disorder, namely, $l_{\rm dis}\ll  l_{\rm e-ph} \ll L$.
For  electron-phonon inelastic scattering  the only conserved quantity is the total density of electron. This implies
\begin{equation}
\label{mu}
\int d\epsilon\,  I_{\rm e-ph}[f]=0\,.
\end{equation}
For systems longer that electron-phonon inelastic length the  distribution function takes the following form
\begin{equation}
\label{f_ph}
f(\epsilon,z)=f_F\left(\frac{\epsilon- \mu(z)}{T} \right),
\end{equation}
where  the temperature  $T$ is equal to the temperature of phonon bath, and the chemical potential satisfies 
\begin{equation}
\left(-\partial_z^2+a^{-1}\partial_z\right)\mu(z)=0 \,.
\end{equation}
For distribution (\ref{f_ph}) Eq.(\ref{s2}) yields
\begin{eqnarray}
S_2=2T\, G(B)\,,
\end{eqnarray}
corresponding to Nyquist noise with phonon-temperature. 
As in normal metals, if current fluctuations are only carried by  fluctuations of chemical potential, they  are not sensitive  to the external bias.

\subsection{Electron-electron relaxation ($L ,a\gg l_{\rm e-e}$)}
\label{el_el}
For a system longer than the electron-electron collision length and with no electron-phonon scattering there are two 
propagating modes: density and temperature fluctuations. The latter arise  due to the energy conservation for electron-electron scattering
\begin{equation}
\label{epsilon}
\int d\epsilon\, \epsilon\, I_{\rm e-e}[f]=0 \,.
\end{equation}
This implies that the mean value of the distribution function has a form
\begin{equation}
\label{f_ee}
f(\epsilon,z)=f_F\left(\frac{\epsilon- \mu(z)}{T(z)} \right).
\end{equation}
Plugin this into Eq.(\ref{Boltzmann}) one finds
 \begin{equation}
\begin{aligned}
& (-\partial_z^2+a^{-1}\partial_z)\,\mu(z)=0 \,,\\
& (-\partial_z^2+a^{-1}\partial_z)\left(\frac{1}{2}\mu^2(z)+\frac{\pi^2}{6} T^2(z)\right)=0 \,. 
\end{aligned}
%\label{}
\end{equation}
Solving these equations one finds the chemical potential
\begin{equation}
\mu(z)=\mu+\frac{eV}{2}
\frac{1+e^{L/a}-2e^{z/a}}{1-e^{L/a}} \,,
\end{equation} 
and the effective temperature
\begin{equation}
T^2(z)=\frac{3e^{2}V^2}{\pi^2}\frac{\left(e^{L/a}-e^{z/a}\right)\left(e^{z/a}-1\right)}{\left(e^{L/a}-1\right)^2}\,,
\end{equation} 
here we assumed $T=0$ at the leads. 
Substituting these expressions into the distribution function, we obtain % ($x=L/a$)
\begin{equation}
S_2=\frac{\sqrt{3}}{8 \pi}\frac{e^3N_{W}\xi V}{L}  \frac{x}{\sinh(x/2)} 
=\frac{\sqrt{3}}{8 \pi}\frac{e^3 N_{W} N_{\phi} V} {\sinh(x/2)} \,,
\end{equation} 
and the corresponding Fano factor 
\begin{equation}
\label{Fee}
F_{\rm ee}(x)=\frac{\sqrt{3}}{4\cosh{x/2}}.
\end{equation}

Note, that unlike normal metal, where different inelastic regimes result in different numerical values of the Fano factors, in Weyl semimetals different inelastic processes  have  different magnetic field  dependence.
For  $L \ll a$,  $S_2\simeq \frac{\sqrt{3}}{4}eI$  as expected for normal metals \cite{Nagaev1995,Kozub1995}. 
In the limit $L \gg a$ the shot noise is exponentially suppressed.
By changing  the magnetic field one should be able to experimentally  explore the crossover from  diffusive to ``ballistic'' regimes.
Moreover, the different parametric dependence in Eqs. (\ref{Fel}) and (\ref{Fee}) allows to estimate $l_{ee}$ from  measurement of shot noise as a function of magnetic field.

% \Dcom{should we calculate  the distribution function that convolutes between elastic and inelastic limits?}
%Note, that unlike normal metal, where different inelastic regimes result in different numerical values of the Fano factors, in Weyl semimetals different inelastic processes  have  different magnetic field  dependence.
\section{Conclusions}
In this paper we  studied the  current noise in Weyl semimetals in the presence of inelastic scattering. We constructed  a Boltzmann-Langevin approach, taking into account chiral anomaly effects. 
Similar to a case  of  elastic propagation \cite{Bagrets,Khalaf2017}, we find that chiral anomaly effects   dominate the evolution of the occupation in phase space  for  samples longer than the 
drift length $a \equiv \xi/N_\phi=   2\pi\nu D/eB$, experimentally controllable by tuning the magnetic field. 
This gives rise to  deterministic (ballistic)  propagation of fluctuations at long samples $ L\gg a$, in a direction determined by the chirality. As a result,  only stochastic processes that occur within a distance $a$ from  the leads,  contribute to the current noise.

%\Dadd{For a closed system,}
We show that inelastic collisions resulting from electron-electron interactions,   modify this  parametric dependence on the magnetic field. By measuring shot noise as a function of  a magnetic field, for different applied voltage, which interpolates between the elastic and inelastic limit, one can estimate the   electron-electron inelastic  length $l_{\rm ee}$.

Finally, we show that  as in normal metals, the presence of electron-phonon scattering suppresses the shot noise, and the current fluctuations correspond to Nyquist noise, governed by the phonon bath temperature.

% \acknowledgements
\section{Acknowledgements}
This work was supported by  ISF (grant 584/14),  Israeli Ministry
of Science, Technology and Space, and   by RFBR grant No 15-52-06009. 
D. M. acknowledges support from the Israel Science Foundation (Grant No. 737/14) and from the European Union's Seventh Framework Programme (FP7/2007- 2013) under Grant No. 631064.
The discussions with D. Bagrets, P. Ostrovsky, A. Stern are gratefully acknowledged. 
 
 %\begin{appendices}
 \appendix
 \section{- Derivation of Eq.(\ref{s2}) }
 \label{derivation}
 \begin{widetext}
 The action (\ref{action}) allows to  compute the correlation functions $D^{\alpha}$ with  $\alpha \in \{R,A,K\}$ 
 In the static limit the retarded propagator
 \begin{equation}
 \bigg[\xi\left(-\partial_z^2+a^{-1}\partial_z\right)+I_{\rm e-e}+I_{\rm e-ph}\bigg]
 D^R(\epsilon,z;\epsilon',z')=\delta(\epsilon-\epsilon')\delta(z-z')\,.
 \end{equation}
The advanced propagator $D^A$ is given by Hermitian conjugation of $D^R$, understood as a matrix, with respect to energy and spatial coordinates.
The Keldysh part of the propagator
\begin{equation}
\label{DK}
D^K(\epsilon,z;\epsilon',z')=-2\xi\int_0^L dz_1\int_{-\infty}^\infty d\epsilon_1D^R(\epsilon,z;\epsilon_1,z_1)\partial_{{z_1}}f(\epsilon_1,z_1)(1-f(\epsilon_1,z_1))\partial_{{z_1}}D^A(\epsilon_1,z_1;\epsilon',z')\,.
\end{equation}
The key observation is that diffusion propagator that enters into Eq.(\ref{gen_noise}) contains diffusion propagators integrated over energy
$\int d\epsilon\,  D^R(\epsilon,z;\epsilon',z')={\bf D^R}(z,z')$ and $\int d\epsilon' D^A(\epsilon,z;\epsilon',z')={\bf D}^A(z,z')$.
It follows from  particle conservation preserved by both electron-electron and electron-phonon scattering, that $\int d\epsilon\, I_{ee} = \int d\epsilon\,    I_{\rm e-ph}=0$.
This implies that propagators integrated over energy  are  equal to those in elastic case. 
Hence the structure of Eq.(\ref{s2}) is preserved in the case of inelastic scattering. We now prove it with   more details in some important limiting cases.

\subsection{$l_{\rm ee},l_{\rm e-ph} \gg L$}
For the system shorter that shortest  inelastic length  the electron-phonon and electron-electron collisions can be neglected. 
In this case the evolution is purely  elastic propagation, and  the energy is conserved.
\begin{equation}
\label{el_diff}
D^\alpha(\epsilon,z;\epsilon',z')=\delta(\epsilon-\epsilon'){\bf D}_\epsilon^\alpha(z,z').
\end{equation}
Here $\alpha \in \{R,A,K\}$ and   we define
\begin{equation}
\label{elastic_diffusion}
{\bf D}^R(z,z')=\frac{a}{\xi}\frac{1}{1-e^{L/a}}
\left\{ 
\begin{array}{ll}
\left(e^{z/a}-1\right)\left(e^{(L-z')/a}-1\right), \,\,\, z<z' \\
\\
\left(e^{z/a}-e^{L/a}\right)\left(e^{-z'/a}-1\right), \,\,\, z>z'
\end{array}
\right. \,,
\end{equation}
The  advanced  propagator  ${\bf D}^A(z,z')$   is  obtained  from ${\bf D}^R(z,z')$  by replacing  $a \rightarrow -a$.
\begin{equation*}
{\bf D}^K_\epsilon(z,z')=-2\xi\int_0^L dz_1\, {\bf D}^R(z,z_1)\partial_{z_1}  f(\epsilon,z_1)(1-f(\epsilon,z_1)\partial _{z_1}{\bf D}^A(z_1,z') 
\end{equation*}
Using  Eq.(\ref{elastic_diffusion}) and integrating over energy and coordinate we come to Eq.\ (\ref{s2}).

\subsection{$l_{\rm e-ph}\ll L$}
The diffusion propagator is
 \begin{equation}
\begin{aligned}
D^R(\epsilon,z;\epsilon',z') & =\partial_\mu  f\left(\frac{\epsilon-\mu(z)}{T}\right) {\bf D}^R(z,z') \, ,\\
D^A(\epsilon,z;\epsilon',z') &={\bf D}^A(z,z') \partial_\mu  f\left(\frac{\epsilon'-\mu(z')}{T}\right) \,, 
\end{aligned}
\label{diff_ph}
\end{equation}
and $D^K$ can be expressed through $D^{R/A}$ via (\ref{DK}). 
Note that   derivatives of distribution function over energies always appear on the outer  part of diffusion propagator. The corresponding integral over energies can be easily performed, 
reproducing Eq.\ (\ref{s2}).

\subsection{$l_{\rm ee} \ll L$  and  $l_{\rm e-ph} \gg L$}
In   case of strong inelastic electron-electron scattering the propagators are 
 \begin{equation}
\begin{aligned}
D^R(\epsilon,z;\epsilon',z')) & =\partial_\mu  f_F\left(\frac{\epsilon-\mu(z)}{T(z)}\right)
\bigg[1+\frac{3}{\pi^2}\frac{\epsilon-\mu(z)}{T(z)}
\frac{\epsilon'-\mu(z')}{T(z')}\bigg]{\bf D}^R(z,z')  \,, \\
D^A(\epsilon,z;\epsilon',z') &=
\bigg[1+\frac{3}{\pi^2}\frac{\epsilon-\mu(z)}{T(z)}
\frac{\epsilon'-\mu(z')}{T(z')}\bigg]{\bf D}^R(z,z')\partial_\mu  f_F\left(\frac{\epsilon'-\mu(z')}{T(z')}\right) \,,
\end{aligned}
\label{diffuson_in}
\end{equation}
The Keldysh part of the propagator is determined via  (\ref{DK}).
Using the idenities
 \begin{equation}
\begin{aligned}
&\int _{-\infty}^\infty \epsilon \frac{\partial f_F(\epsilon)}{\partial \epsilon}\,d\epsilon=0  \,, \\
& \int_{-\infty}^\infty \epsilon  \frac{\partial f_F(\epsilon)}{\partial \epsilon}f_F(\epsilon)(1-f_F(\epsilon))\, d\epsilon=0 \,,
\end{aligned}
%\label{}
\end{equation}
one reduces Eq.(\ref{gen_noise}) to Eq.(\ref{s2}).

\end{widetext}
%  \end{appendices} 

\end{document}